\documentstyle[aps,multicol,eqsecnum,epsf]{revtex}
\begin{document}
\draft
\title{Strongly Temperature Dependent Sliding Friction for a Superconducting 
Interface}
\author{J. B. Sokoloff, M. S. Tomassone and A. Widom}
\address{Physics Department and Center for Interdisciplinary Research on 
Complex Systems, Northeastern University, Boston, MA 02115.}
\date{\today }
\maketitle

\begin{abstract}

A sudden drop in mechanical friction, between an adsorbed nitrogen 
monolayer and a lead substrate, occurs when the lead passes through 
the superconducting transition temperature. We attribute this 
effect to a sudden drop at the superconducting transition temperature 
of the substrate Ohmic heating. The Ohmic heating is due to the 
electronic screening current that results from the sliding 
adsorbed film.
\end{abstract}

\begin{multicols}{2}
\narrowtext

A longstanding question in the field of nanotribology (wearless friction 
between well characterized surfaces) is whether frictional 
dissipation is dominated by the creation of phonons or the creation 
electronic excitations. To study this question, an experiment using a 
quartz crystal microbalance (QCM) was performed in which a monolayer film  
of nitrogen was deposited on flat lead substrate microbalance 
electrodes\cite{dayo} . When the temperature of the lead was lowered below the 
superconducting transition temperature $T_c$, a rapid (nearly discontinuous) 
drop was observed in the monolayer friction. The friction was deduced 
from the damping of the quartz microbalance oscillator. The drop in 
friction was interpreted as a reduction in the substrate electronic 
heating.  

The electronic contribution to sliding friction experienced by a 
non-metallic film sliding on a metallic surface, is usually attributed 
to the creation of excitations of the metallic substrate 
electrons\cite{persson}. However, as the temperature is lowered 
below the substrate 
superconducting transition temperature, the normal (non-superconducting)
fraction of electrons decreases gradually. The friction contribution
due to the creation of electronic excitations might naively be thought
to also decrease relatively gradually. On the basis of the above 
argument, the almost discontinuous experimental results in 
Ref. \cite{dayo} would appear to be surprising. The solution to this
puzzle lies in the nature of the electronic excitations in the 
superconducting substrate.

In addition to the finite frequency electron-hole pair creation 
mechanism of Ref. \cite{persson} for electronic friction, there exists the 
(virtually) zero frequency electron-hole pairs normally considered 
responsible for Ohmic heating in metals. 
Substrate Ohmic heating, induced by a sliding film, will be enhanced 
if the atoms in the film are charged. A net charge may appear 
on an atom if electrons are  donated  to the metallic substrate. 
Such a charge redistribution results in a monolayer surface dipole 
moment. The surface dipole charge structure induces Ohmic 
heating resulting from the electric currents in the metallic substrate 
induced by the sliding monolayer film. The electronic charge redistribution 
follows the film atoms as they move over the substrate surface\cite{tomassone}. 
Unlike the mechanism of Ref. \cite{persson}, this mechanism vanishes very quickly 
as electron pairs form in the superconductor, i.e. practically right at 
$T_c$. We propose this mechanism as the reason for the rapid drop in 
the friction observed in the experiment of Dayo and Krim. 

It is well established that the adsorption of a non-metallic film 
results in an observed change\cite{wang} in the work function (i.e. 
electron chemical potential) of a metallic substrate. This observation 
has been interpreted 
in terms of transfer of electronic charge from the film to the metallic 
substrate, or equivalently as a surface dipole moment per unit area of 
the film atoms. This picture produces the electric fields 
needed for the mechanism in Ref.\cite{tomassone}. 

In Ref. \cite{tomassone}, the force of friction acting on a charged atom whose 
center moves parallel to the substrate a distance Z above its 
surface is defined as the interaction of the atom with 
the electric field due to its image charge. The image electric field is  
$$
{\bf E}_{image}=-q\eta\left(i{\partial\over \partial t}\right)
\left\{{({\bf r}-{\bf R}_i)\over |{\bf r}-{\bf R}_i|^3}\right\}, 
\eqno (1)$$

where the operator $\eta(i\partial/\partial t)$ is constructed by 
the substitution  $\omega$ to $i(\partial/\partial t)$ in the formula 
$\eta(\omega)=
({\epsilon (\omega )-1) /(\epsilon (\omega )+1})$. 
The dielectric constant $\epsilon (\omega )$ is related to 
electrical conductivity $\sigma (\omega )$ via   
$\epsilon (\omega)=1+(4\pi i\sigma (\omega)/\omega)$. 
${\bf R}_i$ is the location of the center of the electrical image of 
the charged atom. If ${\bf R}$=(X,Y,Z) is the location of the center of the 
charged atom, then ${\bf R}_i=(X,Y,-Z)$. The conductivity of the substrate 
can be represented by a Drude model. For the superconducting state,  
$$
\sigma (\omega ) =
\left({n_n e^2\tau_s \over m }\right)
-\left({n_s e^2\over im\omega }\right), 
\eqno (2)$$
where $n_n$ and $n_s$ are the number densities of normal and 
superconducting electrons, respectively, and $\tau_s$ is the scattering 
time for the normal electrons. Since the time Fourier transforms of 
smooth functions of ${\bf R}_i(t)$ are expected to be peaked around 
a frequency of the order of the sliding velocity of the atom divided 
by an atomic length scale (i.e., about $10^{-8}cm$)\cite{dayo}, and since 
the only other frequency in 
the problem (in Guassian units) is determined by 
$\sigma_n= (n_n e^2\tau_s/m)$ (which for $n_s\approx 10^{23}cm^{-3}$ 
and $\tau_s\approx 10^{-14}s^{-1}$ is of the order of $10^{17}s^{-1}$), we are 
justified in expanding the operator $\eta (i\partial/\partial t)$ 
in powers of $i(\partial/\partial t)$. This is 
equivalent to expanding  $\eta (\omega)$ in powers of $\omega $, which 
gives us a series expansion of the friction in powers 
of the sliding velocity. For the superconductor, up to third order   

$$
\eta \left( i{\partial\over \partial t} \right)\approx 
$$ $$
-i\left({ m\over 2\pi n_s e^2}\right )
\left[\left(i{\partial\over \partial t }\right)^2+
\left({n_n\tau_s\over 2\pi e^2 n_s^2}\right)
\left(i{\partial\over \partial t }\right)^3
+ ... \right]. 
\eqno (3)$$

Let us operate with Eq.(3) on the curly bracketed factor in 
Eq. (1), and then set ${\bf r}={\bf R}$. The second derivative 
term gives no contribution to the component of the force parallel 
to the surface of the substrate. The third derivative term gives 
$$
F=\left( {n_n\tau_s V\over n_s h}\right)^2 
\left\{\left({q^2 m\over 2\pi n_n e^2 \tau_s }\right)
\left({{\bf V}\over h^3}\right)\right\}. 
\eqno (4)$$

The terms in curly brackets are the expression for the force of 
friction for the normal state found in Ref.\cite{tomassone}. Hence, 
$\tau_s V/h\approx (10^{-13}sec/10^{-8}cm)(1 cm/sec)=10^{-6}$ 
implies that ${\bf F}$ in the superconducting state will be much 
smaller than the normal state force of friction 
as long as $n_n/n_s<<10^6$. This condition holds true even for 
T very slightly below $T_c$. The density of superconducting 
electrons $n_s$ can be deduced from measurements of the London 
penetration depth $\Lambda=\sqrt{(mc^2)/(4\pi n_s e^2)}$. 

In Ref.\cite{persson}, the ``slip time'' $\tau $ of a film was measured, where 
$\tau$ is defined as the time required for the velocity of a sliding film 
to decay to $1/e$ of its initial value. $\tau^{-1}$ was found 
to decrease by $6.6\times 10^7 sec^{-1}$ on dropping below $T_c$. 
We assume that the lead resistivity sample used in Ref.\cite{dayo} and the lead 
substrate used in the microbalance both have a thickness of $1500 \AA $, 
and that the samples slabs were $3mm \times 3mm$ squares\cite{krim}. 
The resistance for the lead sample given in the experiment was 
$0.04$ Ohms just above $T_c$. Thus the normal state resistivity was 
$6\times 10^{-7}$ Ohm-cm. Equivalently, the conductivity was 
$1.5\times 10^{18}sec^{-1}$ in Gaussian units. On the basis of 
Ref.\cite{persson}, the change in $\tau^{-1}$ at $T_c$ is given by
$$
\Delta (\tau)^{-1}=\left({Z^2 e^2\over 16\pi m\sigma h^3}\right)\approx
$$
$${1^2 (4.8\times 10^{-10} esu)^2 \over 16\pi (4.6\times 10^{-23}gm)(1.5
\times 10^{18}sec^{-1})(10^{-8}cm)^3}, 
\eqno (5)$$

for nitrogen sliding on lead. In Eq.(5), $4.6\times 10^{-23}gm$ is the 
mass of a nitrogen molecule, and $h\approx 10^{-8}cm$ is half of a 
lattice constant for solid nitrogen, which we take as its distance above 
the substrate. Then we obtain $\Delta (\tau)^{-1}=6.6\times 10^7 sec^{-1}$, 
which is consistent with the results of 
Ref.\cite{dayo}
 
Since the picture presented here is based on a model in which the film 
atoms are charged, it is necessary to make some rough estimates of the 
corrugation potential i.e., the dependence of the potential of 
interaction between a film atom on the surface and the ions inside the 
metal substrate on the position of the film atom on the surface.
We wish to determine whether the small values of the corrugation 
needed to explain experiments done with the QCM\cite{tomassone2} can occur
when the film atoms are charged. If the adsorbate atoms are charged, 
it is expected that the main contribution to the corrugation potential 
results from the screened Coulomb interaction between the adsorbate 
atom and the ions. As a rough approximation to this interaction, let us 
use the Thomas-Fermi approximation\cite{ashcroft1} to the interaction between 
an adsorbate atom and an ion in the metal, $(q_1 q_2 e^{-k_s r}/r)$, 
where $q_1$ and $q_2$ are the charges of the two atoms, and $r$ is the 
distance between them. In these estimates, it is assumed that the 
adsorbate atom lies right on the surface and is "bathed" in conduction 
electrons. The inverse screening length 
$k_s\approx [4\pi e^2 g(\epsilon_F)]^{1/2}$, where 
$g(\epsilon_F)$ is the density of states at the Fermi level. Using the 
free electron density of states appropriate for lead\cite{ashcroft2},  
and choosing the effective mass of an electron in lead to be $m=1.3m_e$, 
we obtain $k_s\approx 2.2/\AA $. For these estimates, the  
metal ions near the surface were taken to lie on a triangular lattice 
in which each ion has a charge charge 4e (for lead). We used a lattice 
constant $2.9 \AA $. The interaction between an 
adsorbate atom located $\approx 4 \AA $ above the center of the plane 
along which the centers of the ions of a $900$ atom lattice lie 
is then numerically summed over the positions of the ions in the lattice. 
This interaction was found for three positions of the adsorbate atom: 
(1) above the triangle formed by three near neighbor ions in the lattice, 
(2) above one of the ions, and (3) above the line joining two nearest 
neighbor ions. The potential of interaction was found to be respectively 
$17.6 meV$, $22.3 meV$ and $17.9 meV$. Thus, we conclude that the 
corrugation potential is of the order of a few meV's. This value  
is comparable to that used to successfully simulate QCM measurements 
of friction for rare gas atoms adsorbed on a nobel metal 
substrate.\cite{tomassone2} 

We conclude that the sudden drop in the sliding friction acting on a 
nitrogen film as it slides over a lead substrate, as the lead drops below 
its superconducting transition temperature, can be attributed to 
Ohmic heating induced in the metal by the sliding nitrogen film. 

\acknowledgments

J. B. S wishes to thank the Department of Energy (Grant DE-FG02-96ER45585).

\end{multicols}{2

\end{document}